\pdfoutput=1
\newif\ifpreprint
 \preprinttrue  

\ifpreprint
  \documentclass[11pt,
                 numbers=noenddot,
                 headings=normal,
                 DIV16, BCOR10mm]{scrartcl}
  \usepackage[hyphens]{url}
  \usepackage{fancyhdr}
  \usepackage[linktocpage,breaklinks]{hyperref}
  \usepackage{color}
  \definecolor{comment}{rgb}{0.0,0.6,0.0}
  \usepackage{authblk}
  \usepackage{listings}
  \lstdefinestyle{cppstyle}{language=C++, basicstyle=\ttfamily\small,
    extendedchars=true, escapeinside={/*@}{@*/},
    breaklines=true, breakatwhitespace=true,
    numbers=left, numberstyle=\tiny,
    xleftmargin=5pt,
    keywordstyle=\color{blue}\bfseries,
    stringstyle=\color{red}\ttfamily,
    commentstyle=\color{comment}\ttfamily,
    morecomment=[l][\color{magenta}]{\#}
  }
  \lstnewenvironment{c++}[1][]{\lstset{style=cppstyle}}{}
  \newcommand{\cpp}{\lstinline}
\else
  \documentclass{ansarticle}
\fi

  \usepackage{amsmath,amssymb}
  \usepackage{mathtools}
  \lstdefinestyle{errorstyle}{language=, basicstyle=\ttfamily\small,
    extendedchars=true, escapeinside={/*@}{@*/},
    breaklines=true, breakatwhitespace=true,
    numbers=left, numberstyle=\tiny,
    xleftmargin=5pt,
    keywordstyle=\color{blue}\bfseries,
    stringstyle=\color{red}\ttfamily,
    commentstyle=\color{comment}\ttfamily,
    morecomment=[l][\color{magenta}]{\#}
  }
  \lstnewenvironment{error}[1][]{\lstset{style=errorstyle}}{}

\usepackage[utf8]{inputenc}
\usepackage{overpic}
\usepackage{xspace}
\usepackage{subcaption}
\usepackage{todonotes}
\usepackage{attachfile2}
\usepackage{tikz}

\newcommand{\dune}{\textsc{Dune}\xspace}
\newcommand{\dunemodule}[1]{\texttt{#1}}
\newcommand{\file}[1]{\texttt{#1}}

\graphicspath{{gfx/}}

\title{The interface for functions in the dune-functions module}
\author[1]{Christian Engwer}
\author[2]{Carsten Gräser}
\author[3]{Steffen Müthing}
\author[4]{Oliver Sander}

\affil[1]{Universität Münster, Institute for Computational und Applied Mathematics, christian.engwer@uni-muenster.de}
\affil[2]{Freie Universität Berlin, Institut für Mathematik, graeser@mi.fu-berlin.de}
\affil[3]{Universität Heidelberg, Institut für Wissenschaftliches Rechnen, steffen.muething@iwr.uni-heidelberg.de}
\affil[4]{TU Dresden, Institute for Numerical Mathematics, oliver.sander@tu-dresden.de}
\ifpreprint
\else
  \runningtitle{Functions in dune-functions}
  \runningauthor{C.~Engwer, C.~Gräser, S.~Müthing, O.~Sander}
\fi

\begin{document}

\maketitle

\begin{abstract}
The \dunemodule{dune-functions} \dune module introduces a new programmer interface for discrete and non-discrete functions.
Unlike the previous interfaces considered in the existing \dune modules, it is based on overloading \cpp{operator()},
and returning values by-value.  This makes user code
much more readable, and allows the incorporation of newer C++ features such as lambda expressions.  Run-time polymorphism
is implemented not by inheritance, but by type erasure, generalizing the ideas of the \cpp{std::function} class from the C++11 standard
library.  We describe the new interface, show its possibilities, and measure the performance impact of type erasure and
return-by-value.
\end{abstract}

\section{Introduction}

Ever since its early days, \dune \cite{bastian_et_al:dune1:2008,bastian_et_al:dune2:2008}
has had a programmer interface for functions. This interface was based on the class
\cpp{Function}, which basically looked like
\begin{c++}
template <class Domain, class Range>
class Function
{
  public:
    void evaluate(const Domain& x, Range& y) const;
};
\end{c++}
and is located in the file \file{dune/common/function.hh}.  This class was to serve as a model for duck typing,
i.e., any object with its interface would be called a function. A main feature was that the result of a function
evaluation was {\em not} returned as a return value.
Rather, it was returned using a by-reference argument of the \cpp{evaluate} method.  The motivation for this
design decision was run-time efficiency.  It was believed that returning objects by value would, in practice,
lead to too many unnecessary temporary objects and copying operations.

Unfortunately, the old interface lead to user code that was difficult to read in practice.  For example, to
evaluate the $n$-th Chebycheff polynomial $T(x) = \cos(n\arccos(x))$ using user methods \cpp{my_cos} and
\cpp{my_arccos} would take several lines of code:
\begin{c++}
double tmp1,result;
my_arccos.evaluate(x,tmp1);
my_cos.evaluate(n*tmp1, result);
\end{c++}
Additionally, C++ compilers have implemented return-value optimization (which can return values by-value
without any copying) for a long time, and these implementations have continuously improved in quality.
Today, the speed gain obtained by returning result values in a by-reference argument is therefore not worth
the clumsy syntax anymore (we demonstrate this in Section~\ref{sec:performance_measurements}).
We therefore propose a new interface based on \cpp{operator()} and returning objects by value.  With this
new syntax, the Chebycheff example takes the much more readable form
\begin{c++}
double result = my_cos(n*my_arccos(x));
\end{c++}

To implement dynamic polymorphism, the old functions interface uses virtual inheritance.
There is an abstract base class
\begin{c++}
template <class Domain, class Range>
class VirtualFunction :
  public Function<const Domain&, Range&>
{
public:
  virtual void evaluate(const Domain& x, Range& y) const = 0;
};
\end{c++}
in the file \file{dune/common/function.hh}.  User functions that want to make use of run-time polymorphism
have to derive from this base class.  Calling such a function would incur
a small performance penalty \cite{driesen_hoelzle:1996},
which may possibly be avoided by compiler devirtualization \cite{hubicka:2014}.

The C++ standard library, however, has opted for a different approach.  Instead of deriving different function objects
from a common base class, no inheritance is used at all.  Instead, any object that implements \cpp{operator()} not
returning \cpp{void} qualifies as a function by duck typing.  If a function is passed to another class, the C++ type
of the function is a template parameter of the receiving class.  If the type is not known at compile time, then the
dedicated wrapper class
\begin{c++}
namespace std
{
  template< class Range, class... Args >
  class function<Range(Args...)>;
}
\end{c++}
is used.  This class uses type erasure to allow to handle function objects of different C++ types through a common
interface.  There is a performance penalty for calling functions hidden within a \cpp{std::function}, comparable to
the penalty for calling virtual functions.

The \dunemodule{dune-functions} module \cite{DuneFunctions} picks up these ideas and extends them. The class template
\cpp{std::function}
works nicely for functions that support only pointwise evaluation.  However, in a finite element context, a few
additional features are needed.  First of all, functions frequently need to supply derivatives as well.
Also, for functions defined on a finite element grid, there is typically no single coordinate system of the domain
space.  Function evaluation in global coordinates is possible, but expensive; such functions are usually evaluated
in local coordinates of a given element.  \dunemodule{dune-functions} solves this by introducing \cpp{LocalFunction}
objects, which can be {\em bound} to grid elements.  When bound to a particular element, they behave just like a
\cpp{std::function}, but using the local coordinate system of that element.

The paper is structured as follows.
The main building blocks for the proposed function interfaces, viz.\
callables, concept checks, and type erasure,
are introduced in Section~\ref{sec:interfaces}.
Those techniques are then applied
to implement the extended interfaces
for differentiable functions and grid functions
in Section~\ref{sec:extended_interfaces}. Finally,
we present comparative measurements for the current
and the proposed interface in Section~\ref{sec:performance_measurements}.
The results demonstrate that the increased simplicity, flexibility,
and expressiveness do not have a negative
performance impact.

\section{Building blocks for function interfaces}
\label{sec:interfaces}

\begin{figure}
 \begin{center}
  \includegraphics[width=\textwidth]{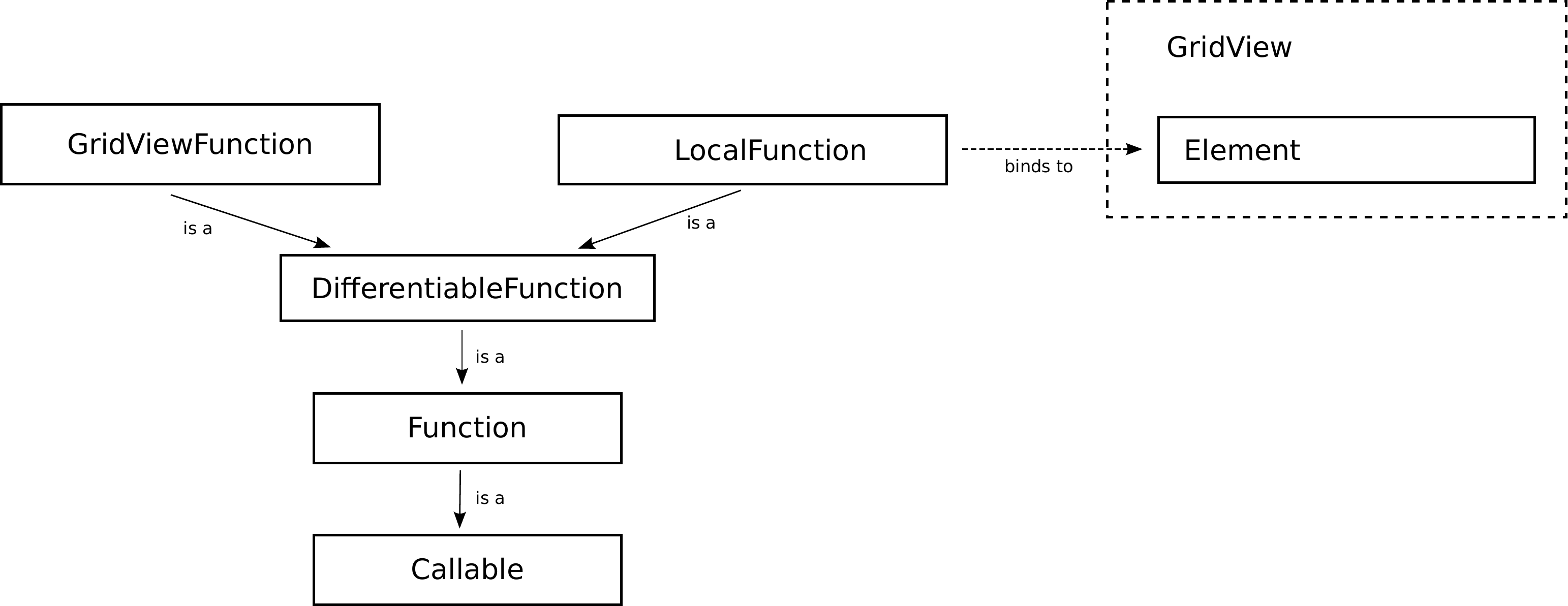}
 \end{center}

 \caption{Diagram of the various function interfaces}
 \label{fig:function_interface_hierarchy}
\end{figure}

The programmer interface for global functions is given as a set of model classes.  These form a conceptual
hierarchy (shown in Figure~\ref{fig:function_interface_hierarchy}), even though no inheritance
is involved at all.  Implementors of the interface need to write classes that have all the methods
and behavior specified by the model classes. This approach results in flexible code.
Concept checking is used to produce readable error messages.
The following sections explain the individual ideas in detail.

\subsection{Callables and functions}
\label{sec:callables_and_functions}

The C++ language proposes a standard way to implement functions.  A language construct is called a
{\em callable} if it provides an \cpp{operator()}.
In the following we will denote a callable as \emph{function}
if that \cpp{operator()} does not return \cpp{void}.
In other words, a function \cpp{foo} is anything that
can appear in an expression of the form
\begin{c++}
auto y = foo(x);
\end{c++}
for an argument \cpp{x} of suitable type. Examples of such constructs are free functions

\begin{minipage}{1.0\linewidth}
\begin{c++}
double sinSquared(double x)
{
  return std::sin(x) * std::sin(x);
}
\end{c++}
\end{minipage}
lambda expressions
\begin{c++}
auto sinSquaredLambda = [](double x){return std::sin(x) * std::sin(x); };
\end{c++}
function objects
\begin{c++}
struct SinSquared
{
  double operator()(double x)
  {
    return std::sin(x) * std::sin(x);
  }
};
\end{c++}
and other things like pointers to member functions, and bind expressions.

  All three examples are callable, i.e., they can be called:
  \begin{c++}
    double a = sinSquared(M_PI);       // free function
    double b = sinSquaredLambda(M_PI); // lambda expression
    SinSquared sinSquaredObject;
    double c = sinSquaredObject(M_PI); // function object
  \end{c++}

Argument and return value do not have to be \cpp{double} at all,
any type is possible.
They can be scalar or vector types, floating point or integer,
and even more exotic data like matrices, tensors, and strings.

To pass a function as an argument to a C++ method, the type of that argument must be a template parameter.
\begin{c++}
template <typename F>
foo(F&& f)
{
  std::cout << "Value of f(42): " << f(42) << std::endl;
}
\end{c++}
Any of the example functions from above can be used as an argument of the method \cpp{foo}:
\begin{c++}
foo(sinSquared);             // call with a free function
foo(sinSquaredLambda);       // call with a lambda expression
foo(sinSquaredObject);       // call with a function object
\end{c++}

\subsection{Concept checks}

The calls to \cpp{F} are fast, because the function calls can be inlined.
On the other hand, it is not clear from the interface of \cpp{foo} what the signature of \cpp{F} should be.
What's worse is that if a callable type with the wrong signature is passed, the compiler error will not occur
at the call to \cpp{foo} but only where \cpp{F} is used, which may be less helpful. To prevent this, \dunemodule{dune-functions}
provides light-weight concept checks for the concepts it introduces. For example, the following alternative
implementation of \cpp{foo} checks whether \cpp{F} is indeed a function in the sense given above
\begin{c++}
template<class F>
void foo(F&& f)
{
  using namespace Dune;
  using namespace Dune::Functions;

  // Get a nice compiler error for inappropriate F
  static_assert(models<Concept::Function<Range(Domain)>, F>(),
      "Type does not model function concept");

  std::cout << "Value of f(42): " << f(42) << std::endl;
}
\end{c++}
If \cpp{foo} is instantiated with a type that is not a function, say, an \cpp{int}, then a
readable error message is produced. For example, for the code
\begin{c++}
foo(1);     // The integer 1 is not a callable
\end{c++}
GCC-4.9.2 prints the error message (file paths and line numbers removed)
\begin{error}
In instantiation of 'void foo(F&&) [with F = int]':
  required from here
error: static assertion failed: Type does not model function concept
   static_assert(models<Function<Range(Domain)>, F>(),
   ^
\end{error}

The provided concept checking facility is based on a list of
expressions that a type must support to model a concept.
The implementation is based on the techniques proposed
by E.~Niebler \cite{Niebler:concepts} and implemented
in the range-v3 library \cite{Niebler:rangev3}.
While the definitions of the function concepts are contained in the
\dunemodule{dune-functions} module, the \cpp{models()} function
that allows to check if a type models the concept is provided by
the module \dunemodule{dune-common}.

\subsection{Type erasure and \texorpdfstring{\cpp{std::function}}{std::function}}

Sometimes, the precise type of a function is not known at compile-time but
selected depending on run-time information. This behavior is commonly referred
to as \emph{dynamic dispatch}.  The classic way to implement this is
virtual inheritance: All functions  must inherit from a virtual base class,
and a pointer to this class is then passed around instead of the function itself.

This approach has a few disadvantages.  For example, all function objects must live on the heap, and a heap
allocation is needed for each function construction.  Secondly, in a derived class, the return value of
\cpp{operator()} must match the return value used in the base class.  However, it is
frequently convenient to also allow return values that are {\em convertible} to the return value of the
base class.  This is not possible in C++. As a third disadvantage,
interfaces can only be implemented intrusively, and having one class implement
more than a single interface is quite complicated.

The C++ standard library has therefore chosen type erasure over virtual inheritance to implement run-time
polymorphism.  Starting with C++11, the standard library contains a class \cite[20.8.11]{cpp_standard:2011}
\begin{c++}
namespace std
{
  template< class Range, class... Args >
  class function<Range(Args...)>
}
\end{c++}
that wraps all functions that map a type \cpp{Domain} to a type (convertible to) \cpp{Range} behind a single
C++ type.
A much simplified implementation looks like the following:
\begin{c++}
template<class Range, class Domain>
struct function<Range(Domain)>
{
  template<class F>
  function(F&& f) :
    f_(new FunctionWrapper<Range<Domain>, F>(f))
  {}

  Range operator() (Domain x) const
  {
    return f_->operator()(x);
  }

  FunctionWrapperBase<Range<Domain>>* f_;
};
\end{c++}
The classes \cpp{FunctionWrapper} and \cpp{FunctionWrapperBase} look like this:
\begin{c++}
template<class Range, class Domain>
struct FunctionWrapperBase<Range(Domain)>
{
  virtual Range operator() (Domain x) const=0;
};

template<class Range, class Domain, class F>
struct FunctionWrapper<Range(Domain), F> :
  public FunctionWrapperBase<Range(Domain)>
{
  FunctionWrapper(const F& f) : f_(f) {}

  virtual Range operator() (Domain x) const
  {
    return f_(x);
  }

  F f_;
};
\end{c++}
Given two types \cpp{Domain} and \cpp{Range}, any function object whose \cpp{operator()} accepts a \cpp{Domain} and
returns something convertible to \cpp{Range} can be stored in a \cpp{std::function<Range(Domain)>}.  For example,
reusing the three implementations of $\sin^2(x)$ from Section~\ref{sec:callables_and_functions}, one can write
\begin{c++}
std::function<double(double)> polymorphicF;
polymorphicF = sinSquared;         // assign a free function
polymorphicF = sinSquaredLambda;   // assign a lambda expression
polymorphicF = SinSquared();       // assign a function object
double a = polymorphicF(0.5*M_PI); // evaluate
\end{c++}
Note how different C++ constructs are all assigned to the same object.  One can even use
\begin{c++}
polymorphicF = [](double x) -> int { return floor(x); };
   // okay: int can be converted to double
\end{c++}
but not
\begin{c++}
polymorphicF = [](double x) -> std::complex<double>
   { return std::complex<double>(x,0); };
   // error: std::complex<double> cannot be converted to double
\end{c++}

Looking at the implementation of \cpp{std::function}, one can see that virtual inheritance is used {\em internally},
but it is completely hidden to the outside. The copy constructor accepts any type as argument.
In a full implementation the same is true for the move constructor,
and the copy and move assignment operators.
For each function type \cpp{F}, an object of type \cpp{FunctionWrapper<Range(Domain),F>} is constructed, which inherits
from the abstract base class \cpp{FunctionWrapperBase<Range(Domain)>}.

Considering the implementation of \cpp{std::function} as described here, one may not expect any run-time gains for
type erasure over virtual inheritance.  While \cpp{std::function} itself does not have any virtual methods, each call
to \cpp{operator()} does get routed through a virtual function.  Additionally, each call to the copy constructor
or assignment operator invokes a heap allocation.  The virtual function call is the price for run-time polymorphism.
It can only be avoided in some cases using smart compiler devirtualization.

To alleviate the cost of the heap allocation, \cpp{std::function} implements a technique called small object optimization.
In addition to the pointer to \cpp{FunctionWrapper}, a \cpp{std::function} stores a small amount of raw memory.
If the function is small enough to fit into this memory, it is stored there.
Only in the case that more memory is needed, a heap allocation is performed.  Small object optimization is therefore
a trade-off between run-time and space requirements.  \cpp{std::function} needs more memory with it, but is faster
{\em for small objects}.

Small object optimization is not restricted to type erasure, and can in principle be used wherever heap allocations
are involved.  However, with a virtual inheritance approach this nontrivial optimization would have to
be exposed to the user code, while all its details are hidden from the user in a type erasure context.

While we rely on \cpp{std::function} as a type erasure class for global functions,
this is not sufficient to represent extended function interfaces as discussed below.
To this end \dunemodule{dune-functions} provides utility functionality to implement
new type erasure classes with minimal effort, hiding, e.g., the details of
small objects optimization. This can be used to implement type erasure
for extended function interfaces that go beyond the ones provided by
\dunemodule{dune-functions}.
Since these utilities are not function-specific, they can
also support the implementation of type-erased
interfaces in other contexts. Similar functionality is, e.g., provided
by the \emph{poly} library (which is part of the \emph{Adobe Source Libraries} \cite{Adobe:asl}),
and the \emph{boost type erasure} library \cite{Watanabe:boost_type_erasure}.

\section{Extended function interfaces}
\label{sec:extended_interfaces}
The techniques discussed until now allow to model functions
\begin{align}\label{eq:function}
    f:\mathcal{D} \to \mathcal{R}
\end{align}
between a domain $\mathcal{D}$ and a range $\mathcal{R}$
by interfaces using either static or dynamic dispatch.
In addition to this, numerical applications often require to model
further properties of a function like differentiability or the fact that
it is naturally defined locally on grid elements. In this section
we describe how this is achieved in the \dunemodule{dune-functions} module
using the techniques described above.

\subsection{Differentiable functions}
\label{sec:differentiable_functions}

The extension of the concept for a function \eqref{eq:function} to a differentiable
function requires to also provide access to its derivative
\begin{align}
    Df : \mathcal{D} \to L(\mathcal{D}, \mathcal{R})
\end{align}
where, in the simplest case, $L(\mathcal{D}, \mathcal{R})$ is the set of linear maps from the affine hull of $\mathcal{D}$ to $\mathcal{R}$.
To do this, the \dunemodule{dune-functions} module extends the ideas from the previous section
in a natural way.  A C++ construct is a differentiable function
if, in addition to having \cpp{operator()} as described above, there is a free method \cpp{derivative} that returns a function
that implements the derivative.  The typical way to do this will be a friend member method as illustrated in the class template
\cpp{Polynomial}:
\begin{c++}
template<class K>
class Polynomial
{
public:

  Polynomial() = default;
  Polynomial(const Polynomial& other) = default;
  Polynomial(Polynomial&& other) = default;

  Polynomial(std::initializer_list<double> coefficients) :
    coefficients_(coefficients) {}

  Polynomial(std::vector<K>&& coefficients) :
      coefficients_(std::move(coefficients)) {}

  Polynomial(const std::vector<K>& coefficients) :
    coefficients_(coefficients) {}

  const std::vector<K>& coefficients() const
  { return coefficients_; }

  K operator() (const K& x) const
  {
    auto y = K(0);
    for (size_t i=0; i<coefficients_.size(); ++i)
      y += coefficients_[i] * std::pow(x, i);
    return y;
  }

  friend Polynomial derivative(const Polynomial& p)
  {
    std::vector<K> dpCoefficients(p.coefficients().size()-1);
    for (size_t i=1; i<p.coefficients_.size(); ++i)
      dpCoefficients[i-1] = p.coefficients()[i]*i;
    return Polynomial(std::move(dpCoefficients));
  }

private:
  std::vector<K> coefficients_;
};
\end{c++}

To use this class, write
\begin{c++}
auto f = Polynomial<double>({1, 2, 3});
double a = f(0.5*M_PI);
double b = derivative(f)(0.5*M_PI);
\end{c++}
Note, however, that the \cpp{derivative} method may be expensive, because it needs to compute the entire derivative function.
It is therefore usually preferable to call it only once and to store the derivative function separately.
\begin{c++}
auto df = derivative(f);
double b = df(0.5*M_PI);
\end{c++}
Functions supporting these operations are described by \cpp{Concept::DifferentiableFunction},
provided in the \dunemodule{dune-functions} module.

When combining differentiable functions and dynamic polymorphism, the \cpp{std::function} class cannot be used as is,
because it does not provide access to the \cpp{derivative} method.  However, it can serve as inspiration for more general
type erasure wrappers. The \dunemodule{dune-functions} module provides the class
\begin{c++}
template<class Signature,
         template<class> class DerivativeTraits=DefaultDerivativeTraits,
         size_t bufferSize=56>
class DifferentiableFunction;
\end{c++}
in the file \file{dune/functions/common/differentiablefunction.hh}. Partially, it is a reimplementation of
\cpp{std::function}. The first template argument of \cpp{DifferentiableFunction} is equivalent to the template
argument \cpp{Range(Args...)} of \cpp{std::function}, and \cpp{DifferentiableFunction} implements a method
\begin{c++}
Range operator() (const Domain& x) const;
\end{c++}
This method works essentially like the one in \cpp{std::function}, despite the fact that its argument type
is fixed to \cpp{const Domain&} instead of \cpp{Domain}, because arguments of a ``mathematical function''
are always immutable. 
Besides this, it also implements a free method
\begin{c++}
friend DerivativeInterface derivative(const DifferentiableFunction& t);
\end{c++}
that wraps the corresponding method of the function implementation.
It allows to call the \cpp{derivative} method for objects whose precise type is determined only
at run-time:
\begin{c++}
DifferentiableFunction<double(double)> polymorphicF;
polymorphicF = Polynomial<double>({1, 2, 3});
double a = polymorphicF(0.5*M_PI);
auto polymorphicDF = derivative(polymorphicF);
double b = polymorphicDF(0.5*M_PI);
\end{c++}

While the domain of a derivative is $\mathcal{D}$, the same as the one of the
original function, its range is $L(\mathcal{D},\mathcal{R})$.
Unfortunately, it is not feasible to always infer the best C++ type for objects from $L(\mathcal{D},\mathcal{R})$.
To deal with this, \dunemodule{dune-functions} offers the
\cpp{DerivativeTraits} mechanism that maps the signature
of a function to the range type of its derivative.  The line
\begin{c++}
using DerivativeRange = DerivativeTraits<Range(Domain)>::Range;
\end{c++}
shows how to access the type that should be used
to represent elements of $L(\mathcal{D},\mathcal{R})$.
The template \cpp{DefaultDerivativeTraits} is specialized
for common combinations of \dune matrix and vector types, and provides
reasonable defaults for the derivative
ranges. However, it is also possible to change
this by passing a custom \cpp{DerivativeTraits} template to the
interface classes, e.g., to allow optimized application-specific
matrix and vector types or use suitable representations for
other or generalized derivative concepts.

Currently the design of \cpp{DifferentiableFunction} differs from
\cpp{std:function} in that it only considers a single argument, but
this can be vector valued.

\subsection{GridView functions and local functions}

A very important class of functions in any finite element application are discrete functions, i.e., functions that
are defined piecewisely with respect to a given grid.  Such functions are typically too expensive to evaluate in
global coordinates. Luckily this is hardly ever necessary. Instead, the arguments for such functions are a grid
element together with local coordinates of that element.
Formally, this means that we have localized versions
\begin{align*}
    f_e = f \circ \Phi_e : \hat{e} \to \mathcal{R},
    \qquad
    \text{$e$ is element of the grid},
\end{align*}
of $f$, where $\Phi_e:\hat{e} \to e$ is a parametrization of a \emph{grid element} $e \subset \mathcal{D}$
over a reference element $\hat{e}$.

To support this kind of function evaluation, \dune has provided interfaces in the style of
\begin{c++}
void evaluateLocal(Codim<0>::Entity element,
                   const LocalCoordinates& x,
                   Range& y);
\end{c++}
Given an element \cpp{element} and a local coordinate \cpp{x}, such a method would evaluate the function at
the given position, and return the result in the third argument \cpp{y}.
This approach is currently used, e.g., in the
grid function interfaces of the discretization modules \dunemodule{dune-pdelab} and \dunemodule{dune-fufem}.

There are several disadvantages to this approach. First, we have argued earlier that return-by-value is preferable to
return-by-reference.  Hence, an obvious improvement would be to use
\begin{c++}
Range operator()(Codim<0>::Entity element, cost LocalCoordinates& x);
\end{c++}
instead of the \cpp{evaluateLocal} method.
However, there is a second disadvantage.  In a typical access pattern in a finite element implementation,
a function evaluation on a given element is likely to be followed by evaluations on the same element.
For example, think of a quadrature loop that evaluates a coefficient function at all quadrature points
of a given element.  Function evaluation in local coordinates of an element can involve some setup code
that depends on the element but not on the local coordinate~\cpp{x}. This could be, e.g., pre-fetching of those
coefficient vector entries that are needed to evaluate a finite element function on the given element,
or retrieving the associated shape functions.

In the approaches described so far in this section, this setup code is executed again and again for each evaluation
on the same element. To avoid this we propose the following usage pattern instead:
\begin{c++}
auto localF = localFunction(f);
localF.bind(element);
auto y = localF(xLocal);           // evaluate f in element-local coordinates
\end{c++}
Here we first obtain a \emph{local function}, which represents the restriction of \cpp{f} to a single
element. This function is then bound to a specific element using the method
\begin{c++}
void bind(Codim<0>::Entity element);
\end{c++}
This is the place for the function to perform any required setup procedures.
Afterwards the local function can be evaluated using the interface described above,
but now using local coordinates with respect to the element that the local function is bound to. The same
localized function object can be used for other elements by calling \cpp{bind} with a different argument.
Functions supporting these operations are called {\em grid view functions}, and described by \cpp{Concept::GridViewFunction}
The local functions are described by \cpp{Concept::LocalFunction}.
Both concepts are provided in the \dunemodule{dune-functions} module.

Since functions in a finite element context are usually at least piecewise
differentiable, grid view functions as well as local functions provid the full
interface of differentiable functions as outlined in Section~\ref{sec:differentiable_functions}.
To completely grasp the semantics of the interface, observe that strictly speaking localization does
not commute with taking the derivative. Formally, a localized version of the derivative is given by
\begin{align}
\label{eq:localized_derivative}
    (Df)_e : \hat{e} \to L(\mathcal{D}, \mathcal{R}), \qquad (Df)_e = (Df) \circ \Phi_e.
\end{align}
In contrast, the derivative of a localized function is given by
\begin{align*}
    D(f_e) : \hat{e} \to L(\hat{e}, \mathcal{R}), \qquad D(f_e) = ((Df) \circ \Phi_e) \cdot D \Phi_e.
\end{align*}
However, in the \dunemodule{dune-functions} implementation, the derivative of a local function
does by convention always return values {\em in global coordinates}. Hence, the functions
\cpp{dfe1} and \cpp{dfe2} obtained by
\begin{c++}
auto de = derivative(f);
auto dfe1 = localFunction(df);
dfe1.bind(element);

auto fe = localFunction(f);
fe.bind(element);
auto dfe2 = derivative(fe);
\end{c++}
both {\em behave the same}, implementing $(Df)_e$ as in~\eqref{eq:localized_derivative}.
This is motivated by the fact that $D(f_e)$ is
hardly ever used in applications, whereas $(Df)_e$ is needed
frequently. To express this mild inconsistency in the interface,
a local function uses a special \cpp{DerivativeTraits} implementation
that forwards the derivative range to the one of the corresponding
global function.

Again, type erasure classes allow to use grid view and local functions in a polymorphic way.
The class
\begin{c++}
template<class Signature,
         class GridView,
         template<class> class DerivativeTraits=DefaultDerivativeTraits,
         size_t bufferSize=56>
class GridViewFunction;
\end{c++}
stores any function that models the \cpp{GridViewFunction}
concept with given signature and grid view type.
Similarly, functions modeling the \cpp{LocalFunction} concept
can be stored in the class
\begin{c++}
template<class Signature,
         class Element,
         template<class> class DerivativeTraits=DefaultDerivativeTraits,
         size_t bufferSize=56>
class LocalFunction;
\end{c++}
These type erasure classes can be used in combination:
\begin{c++}
GridViewFunction<double(GlobalCoordinate), GridView> polymorphicF;
polymorphicF = f;
auto polymorphicLocalF = localFunction(polymorphicF);
polymorphicLocalF.bind(element);
LocalCoordinate xLocal = ... ;
auto y = polymorphicLocalF(xLocal);
\end{c++}
Notice that, as described above, the \cpp{DerivativeTraits} used
in \cpp{polymorphicLocalF} are not the same as the ones used
by \cpp{polymorphicF}.  Instead, they are a special implementation forwarding
to the global derivative range even for the domain type \cpp{LocalCoordinate}.

\section{Performance measurements}
\label{sec:performance_measurements}

In this last chapter we investigate how the interface design for functions in \dune influence the run-time efficiency.
Two particular design choices are expected to be critical regarding execution speed:
(i)~returning the results of function evaluations
by value involves temporary objects and copying unless the compiler is smart enough to remove those using
return-value-optimization.  In the old interface, such copying could
not occur by construction, (ii)~using type erasure instead of
virtual inheritance for dynamic polymorphism. While there are fewer reasons to
believe that this may cause changes in execution time, it is still worthwhile to check empirically.

As a benchmark we have implemented a small C++ program that computes the integral
\begin{equation*}
    I(f) \coloneqq \int_0^{1} f(x)\,dx
\end{equation*}
for different integrands, using a standard composite mid-point rule.
We chose this problem because it is very simple, but still an actual numerical algorithm.  More importantly,
most of the time is spent evaluating the integrand function. Finally, hardly any main memory is needed,
and hence memory bandwidth limitations will not influence the measurements.

The example code is a pure C++11 implementation with no reference to \dune.  The relevant interfaces from \dune
are so short that it was considered preferable to copy them into the benchmark code to allow easier building.
The code is available in a single file attached to this pdf document, via the icon in the margin.%
\marginpar{\attachfile[author={C. Engwer, C. Gräser, S. Müthing, and O. Sander},
                            color = 1 0 0,
                            mimetype=text/plain,
                            description=Benchmark code for the functions interface]{src/integration-test.cc}}

To check the influence of return types with different size we used integrands of the form
\begin{align*}
    f:\mathbb{R} \to \mathbb{R}^N, \qquad f(x)_i = x+i-1, \qquad i=1,\dots,N,
\end{align*}
for various sizes $N$. This special choice was made to keep the computational work done
inside of the function to a minimum while avoiding compiler optimizations that replace the
function call by a compile-time expression.
The test was performed with $n= \lfloor 10^8/N\rfloor$ subintervals for the composite mid-point
rule leading to $n$ function evaluations, such that the timings are directly comparable for different
values of $N$.

For the test we implemented four variants of function evaluation:
\begin{enumerate}
    \item[(a)]
        Return-by-value with static dispatch using plain \cpp{operator()},
    \item[(b)]
        Return via reference with static dispatch using \cpp{evaluate()},
    \item[(c)]
        Return-by-value with dynamic dispatch using \cpp{std::function::operator()},
    \item[(d)]
        Return via reference with dynamic dispatch using \cpp{VirtualFunction::evaluate()},
        as in the introduction.
\end{enumerate}

The test was performed with $N=1,\dots,16$ components for the function range,
using \cpp{double} to implement the components.
We used GCC-4.9.2 and Clang-3.6 as compilers, as provided by the Linux
distribution Ubuntu 15.04. To avoid cache effects and to eliminate outliers we did a warm-up
run before each measured test run and selected the minimum of four subsequent
runs for all presented values.

\begin{figure}
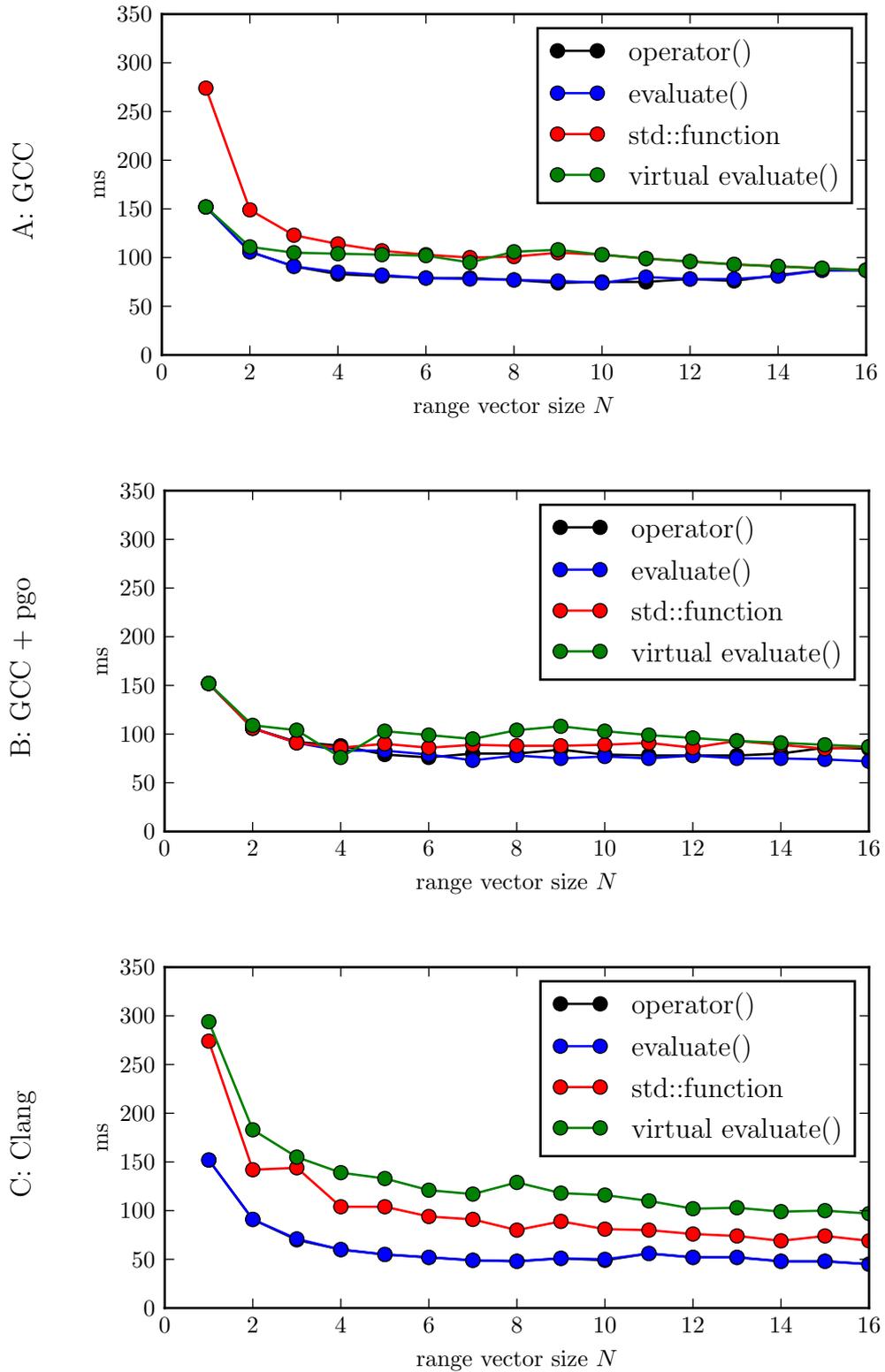

    \centering%
    \large
    \raisebox{-0.5\height+2ex}{
      \rotatebox{90}{A: GCC}}
    \raisebox{-0.5\height}{
      \input{gfx/timings_gcc.pgf}}\\
    \raisebox{-0.5\height+2ex}{
      \rotatebox{90}{B: GCC + pgo}}
    \raisebox{-0.5\height}{
      \input{gfx/timings_gcc_pgo.pgf}}\\
    \raisebox{-0.5\height+2ex}{
      \rotatebox{90}{C: Clang}}
    \raisebox{-0.5\height}{
      \input{gfx/timings_clang.pgf}}%
    \caption{Timings for $\lfloor 10^8/N \rfloor$ function calls over
      varying vector size $N$ using (A) GCC, (B) GCC with profile-guided optimization, and
      (C) Clang.}%
    \label{fig:timings}
\end{figure}

Figure~\ref{fig:timings}.A
shows the execution time in milliseconds over $N$
when compiling with GCC-4.9 and the compiler options~{\tt -std=c++11 -O3 -funroll-loops}.
One can observe that the execution time is the same for variants (a) and (b) and all values
of $N$. We conclude that for static dispatch there is no run-time overhead when using
return-by-value, or, more precisely, that the compiler is able to optimize away any overhead.
Comparing the dynamic dispatch variants (c) and (d) we see that for small values of $N$ there is an overhead
for return-by-value with type erasure compared to classic virtual inheritance.
This is somewhat surprising since pure return-by-value does not impose
an overhead, and dynamic dispatch happens for both variants.

Guessing that the compiler is not able to optimize the nested function
calls in the type erasure interface class \cpp{std::function} to full extent,
we repeated the tests using \emph{profile guided optimization}.
To this end the code was first compiled using the additional option
{\tt -fprofile-generate}. When running the obtained program once,
it generates statistics on method calls that are used by
subsequent compilations with the additional option
{\tt -fprofile-use} to guide the optimizer.
The results depicted in Figure~\ref{fig:timings}.B show that
the compiler is now able to generate code that performs equally
well for variant (c) and (d). In fact variant (c) is sometime even
slightly faster.

Finally, Figure~\ref{fig:timings}.C shows results for
Clang-3.6 and the compiler options
{\tt -std=c++11 -O3 -funroll-loops}. Again variants (a) and (b) show identical results.
In contrast, variant (c) using \cpp{std::function} is now clearly superior compared
to variant (d).
Note that we only used general-purpose optimization options and
that this result did not require fine-tuning with more
specialized compiler flags.

\section{Conclusion}
We have presented a new interface for functions in \dune, which is implemented in the new
\dunemodule{dune-functions} module. The interface follows the
ideas of callables and \cpp{std::function} from the C++ standard library, and generalises these concepts to allow for
differentiable functions and discrete grid functions.
For run-time polymorphism we offer corresponding type erasure classes
similar to \cpp{std::function}. The performance of these new
interfaces was compared to existing interfaces in \dune.
When using the optimization features of modern compilers, the proposed new interfaces
are at least as efficient as the old ones, while being much easier to read and use.

\bibliographystyle{plain}
\bibliography{engwer_graeser_muething_sander_functions}
\end{document}